\documentclass[twocolumn,preprintnumbers,amsmath,amssymb,prb,aps]{revtex4}
\usepackage{amsmath}
\usepackage{amsfonts}
\usepackage{amssymb}

\usepackage{epsfig}
\usepackage{graphicx}
\usepackage{mhchem}
\usepackage{xcolor}
\usepackage{dcolumn}
\usepackage{bm} 
\usepackage{epstopdf}
\usepackage{amsmath}

\begin{document}

\title{Lattice effects on the physical properties of half doped perovskite ruthenates}
\author{Jaskirat Brar$^1$}
\author{Saurabh Singh$^{2,3}$}
\altaffiliation{Present address: Pennsylvania State University}
\author{Kentaro Kuga$^{2,3}$}
\author{Priyamedha Sharma$^1$}
\author{Bharath M$^1$}
\author{Tsunehiro Takeuchi$^2$$^,$$^3$}
\author{R. Bindu$^1$}
\affiliation{$^1$School of Physical Sciences, Indian Institute of Technology Mandi, Kamand, Himachal Pradesh- 175005, India}
\affiliation{$^2$Toyota Technological Institute, Nagoya, Aichi 468-8511, Japan}
\affiliation{$^3$Japan Science and Technology Agency, Kawaguchi, Saitama 332-0012, Japan}

\date{\today}
\begin{abstract}
We investigate the unusual phase transitions in SrRuO$_{3}$ and Sr$_{0.5}$Ca$_{0.5}$Ru$_{1-x}$Cr$_{x}$O$_{3}$ (x=0,0.05 and 0.1) employing x-ray diffraction, resistivity, magnetic studies and x-ray photoemission spectroscopy. Our results show the compounds undergo crossover from $\emph{itinerant}$ ferromagnetism to $\emph{localised}$ ferromagnetism. The combined studies suggests Ru and Cr to be in 4+ valence state. A Griffith phase and an enhancement in Curie temperature (Tc) from 38 K to 107 K is observed with Cr doping. A shift in the chemical potential towards the valence band is observed with Cr doping. In the metallic samples, interestingly, a direct link between the resistivity and orthorhombic strain is observed. Detailed studies in this direction will be helpful to understand the nature of interactions and hence manoeuvre its properties. In the non metallic samples, the resistivity is mainly governed by disorder and electron-electron correlation effects. The value of the resistivity for the 5\% Cr doped sample suggests semi metallic behaviour. Understanding its nature in detail using electron spectroscopic techniques could unravel the possibility of its utility in high mobility transistors at room temperature and its combined property with ferromagnetism will be helpful in making spintronic devices.								

\end{abstract}

\pacs{61.05.cp,75.47,71.27.+a,73.43.Qt}

\maketitle

\section{Introduction}

SrRuO$_{3}$ is an infinite layer material in the Ruddlesden popper series\cite{koster2012}. This compound is a Fermi liquid at low temperatures\cite{capogna2002}, ferromagnetic below 165 K\cite{longo1968}, shows invar effect\cite{kiyama1996} and is a bad metal\cite{emery1995} at high temperatures. On Ca doping at the Sr site\cite{singh2007}, throughout the series, the structure is orthorhombic but with increased distortion. The end compound, CaRuO$_{3}$ is a paramagnet and exhibits non Fermi liquid behaviour at low temperatures\cite{singh2007}. The magnetic anisotropy is also unusually large in the end compounds\cite{longo1968}. Sr$_{1-y}$Ca$_{y}$RuO$_{3}$ exhibits a smeared quantum phase transition in the range y = 0.6 to 0.8 in bulk\cite{cao1997,demko2012} thereby making it as a test ground for studying quantum phenomena. It is also important to note that the critical concentration of Ca at which the T$_{c}$ goes to zero also depends on the sample preparation, experimental protocol etc\cite{demko2012}. These compounds are widely used in spintronics\cite{gu2022} to make superlattice of oxides and is also a good candidate for perpendicular magnetic anisotropy based spintronic material. In these compounds, reports show that structure plays an important role in driving the magnetic ground state\cite{singh2007}. Interestingly, in these ruthenates, on Cr doping at Ru site, an increase in the Tc has been observed\cite{pi2002,dabrowski2005}. But by doping other 3d transition metals, a decrement in the Tc occurs\cite{pi2002,tong2011}. Such behaviours are puzzling.

\begin{figure}[h!]
\includegraphics [width=0.98\linewidth]{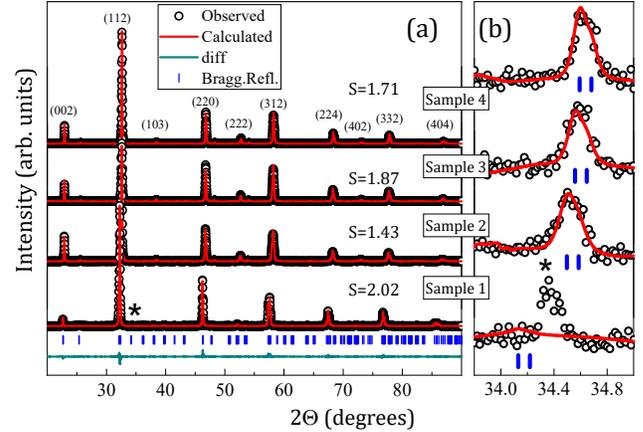}
\vspace{-37ex}
\caption{(a)Room temperature Rietveld refinement of xrd patterns of the samples under study. The open circles and solid lines represent the observed and the calculated patterns, respectively. The ticks represent the Bragg reflections; (b)The star symbol marks the impurity feature observed in sample 1 only.}
\label{Rietveld}
\vspace{-2ex}
\end{figure}

There are various contradicting reports on the origin of the enhancement in the Tc for Cr doped \ce{SrRuO_{3}}. The increase in Tc has been explained on the basis of the minority band double exchange interaction involving Cr$^{3+/4+}$ and Ru$^{4+/5+}$ ions\cite{dabrowski2005,zhang2011}. To clarify this mechanism, information about the valence state of Ru and Cr ions is crucial. Zhang et. al.\cite{zhang2011} have reported that the ionic state of Chromium is Cr$^{3+}$ in Cr doped \ce{SrRuO_{3}} and the presence of Ru$^{5+}$ ionic state is essential in increasing T$_{C}$. However, the ionic states of Cr and Ru are not clear in this system and an ionic state of Cr$^{4+}$ has also been reported in literature\cite{durairaj2006}. The changes in the structural parameter values and the effective magnetic moment values with change in Cr concentration supports the presence of Cr$^{4+}$ ionic state\cite{dabrowski2005,durairaj2006}. The Cr$^{4+}$ ionic state has only 2 electrons in the t$_{2g}$ orbitals which means that the hybridization with Ru$^{4+}$ state having 4 electrons in the t$_{2g}$ state is weaker. This results in a narrow band near the fermi edge, giving rise to stronger magnetism as per the

\begin{figure}[h!]
 \vspace{-5ex}
\includegraphics[width=1.7\linewidth]{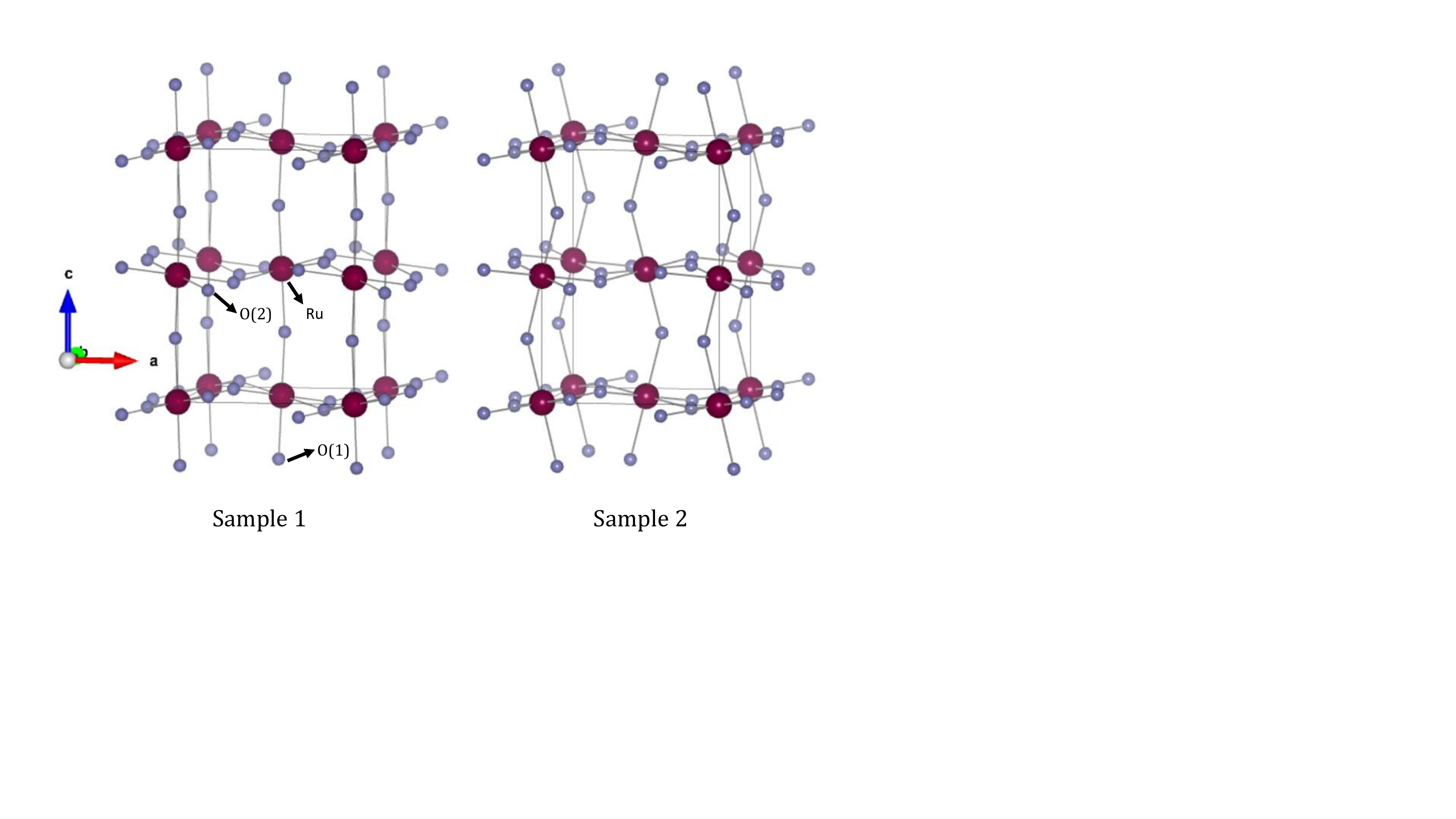}
 \vspace{-25ex}
\caption{RT crystal structure of samples 1 and 2 depicting the extent of the tilt in the octahedra. In this structure, only oxygen ions occupy two sites labelled as O(1) (apical) and O(2)(basal).}
\label{structure}
\vspace{-3ex}
\end{figure}

Stoner criteria\cite{durairaj2006}. We have chosen to study the effect of 5 and 10\% Cr doping in Sr$_{0.5}$Ca$_{0.5}$RuO$_{3}$ considering the fact that this half doped system lies close to the quantum phase transition. The radial extent of 3d states in Cr ion is smaller than the 4d states in Ru ion and also with Ca doping the 4d states are expected to become localised. In such situation, a small change in the structural parameters in terms of the tilt or rotation of the RuO$_{6}$ octahedra can bring about multiple phase transitions. Such localisation effects are expected to make the role of electron-electron interaction, electron phonon coupling, magnetic anisotropy and disorder to be significant.

Keeping all this in mind, we have undertaken x-ray diffraction, magnetic, transport and x-ray photoemission studies to know the valence state of the Ru and Cr ion, origin of the phase transition as a function of composition and temperature and the behaviour of Tc in SrRuO$_{3}$ to Sr$_{0.5}$Ca$_{0.5}$Ru$_{1-x}$Cr$_{x}$O$_{3}$ (x=0,0.05 and 0.1). Our results show a transition from itinerant ferromagnetism to localized ferromagnetism with change in composition. The Cr doped samples show an increase in the Tc and the doped compounds stabilizes Griffith phase. The samples that exhibit metallicity show direct connection between the resistivity and orthorhombic strain but for the rest of the samples the transport is mainly governed by disorder and electron electron correlation effects. Our results show that Ru and Cr stabilises in 4+ valence state. With respect to undoped sample, the Cr doped compound shows a shift in the chemical potential towards the valence band. Understanding the non-metallic nature of the 5\% Cr doped compounds may find its applicability in electronic and spintronic devices.

\section{Experimental}

Polycrystalline samples \ce{SrRuO_{3}}, \ce{Ca_{0.5}Sr_{0.5}RuO_{3}}, \ce{Ca_{0.5}Sr_{0.5}Ru_{0.95}Cr_{0.05}O_{3}} and \ce{Ca_{0.5}Sr_{0.5}Ru_{0.9}Cr_{0.1}O_{3}} were prepared by the solid state reaction method by using \ce{SrCO_{3}}($99.995\%$), \ce{CaCO_{3}}($99.995\%$), \ce{RuO_{2}}($99.9\%$) and \ce{Cr_{2}O_{3}}($99.9\%$) raw materials supplied by Sigma Aldrich. \ce{RuO_{2}} was preheated at 400$^{\circ}$C for 8 hours to remove any moisture. Stoichiometric amounts of raw materials were mixed and ground thoroughly for 4 hours and calcined at 1000$^{\circ}$C for 24 hours. The calcined samples were ground and pressed into pellets of 10 mm diameter and then sintered at 1250$^{\circ}$C for 72 hours with two intermediate grindings. The heating and cooling rates were fixed at 5$^{\circ}$C/min during the sintering process.

Temperature dependent x-ray diffraction measurements (xrd) were done using Rigaku Smartlab x-ray diffractometer for temperature range 300 K to 10 K. Temperature dependent dc susceptibility measurements for all the samples except \ce{Ca_{0.5}Sr_{0.5}Ru_{0.95}Cr_{0.05}O_{3}} were carried out using Quantum design SQUID magnetometer for temperature range 300 K to 1.8 K. For the case of \ce{Ca_{0.5}Sr_{0.5}Ru_{0.95}Cr_{0.05}O_{3}}, the measurements were done using vibrating sample magnetometer (VSM). The field dependent magnetisation measurements upto $5T$ were carried out using Quantum design SQUID magnetometer.

The temperature dependent resistivity measurements were carried out using four probe method, commercial physical properties measurement system from Quantum design, USA.

The room temperature x-ray photoemission spectra (xps) were collected using a monochromatic AlK$\alpha$ source and R3000 Scienta analyser having energy resolution of 400meV. All the core levels were collected after surface cleaning the samples (pellet) in situ by scraping with a diamond file in the chamber with base pressure of $\sim$ $2\times10^{-10}$mbar. In the forthcoming sections, we have labelled the samples \ce{SrRuO_{3}}, \ce{Sr_{0.5}Ca_{0.5}RuO_{3}}, \ce{Sr_{0.5}Ca_{0.5}Ru_{0.95}Cr_{0.05}O_{3}} and \ce{Sr_{0.5}Ca_{0.5}Ru_{0.9}Cr_{0.1}O_{3}} as sample 1, 2, 3 and 4, respectively.

\section{Computational details}
Density of states(DOS) calculations were performed on \ce{SrRuO_{3}} using full potential linearized augmented plane wave method(FPLAPW) as implemented in the Elk code\cite{elk}. The PBEsol exchange correlation functional\cite{perdew2008} was used. The crystal structure parameters obtained after Rietveld refinement of the xrd patterns collected at room temperature were used for calculations. The muffin-tin radii for Sr, Ru and O atoms were considered to be 2.4, 2.0 and 1.6 Bohr. The difference in total energy required for the termination of the self consistent loop was set to be less than $10^{-4}$ Hartree/cell.

\section{Results and Discussions}

Fig.\ref{Rietveld} shows the Rietveld refinement of the room temperature(RT) xrd patterns of samples 1-4. All the compounds stabilize in orthorhombic structure with \textit{Pbnm} space group. In this space group, the wyckoff positions of Sr is 4c (x,y,1/4), Ru is 4a (0,0,0), O1 is 4c and O2 is 8d (x,y,z). The crystal structure is shown in Fig.\ref{structure}.The \ce{RuO_{6}} octahedra show a tilt along the c axis and a rotation in the ab plane in case of all the compounds studied, Fig.\ref{structure}. This is in line with the literature\cite{kennedy2002,bushmeleva2006}. The features corresponding to the tilting of \ce{RuO_{6}} octahedra\cite{kennedy2002,glazer1975} were found in the xrd patterns. The lattice parameter values thus obtained for the parent compound \ce{SrRuO_{3}} are in line with the literature\cite{kennedy2002}. An unidentified impurity peak\cite{singh2007,rao2001} is observed only in case of the parent compound $\sim$ 34.4$^{\circ}$ marked as $^{\ast}$, Fig.\ref{Rietveld}.

\subsection{Magnetic studies}

\begin{figure}[h!]
\includegraphics[width=0.98\linewidth]{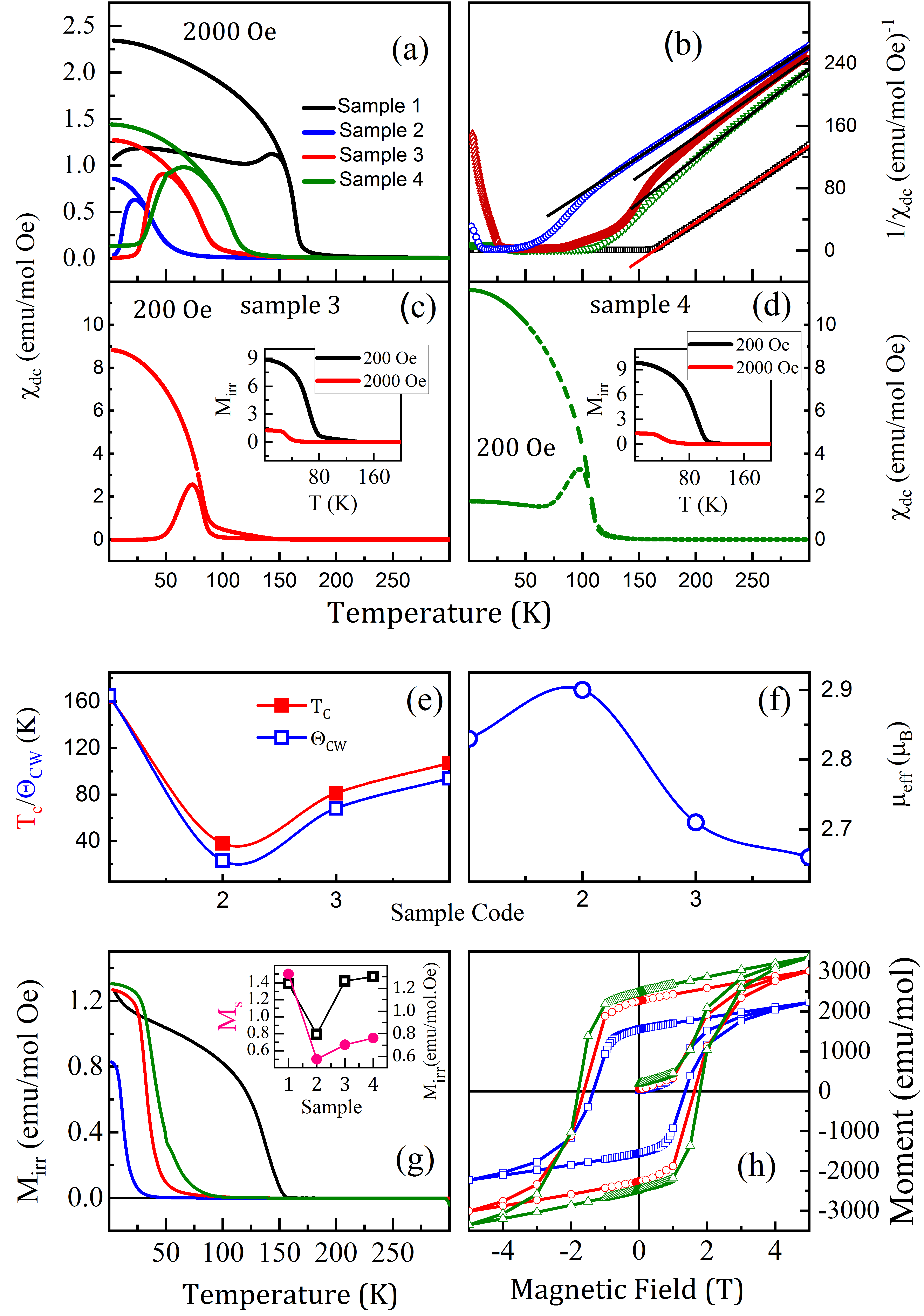}
\caption{(a)FC and ZFC dc susceptibility of all compositions under study. (b)Inverse susceptibility as a function of temperature(Solid lines show the Curie-Weiss fitting). (c) and (d) show the dc susceptibility for samples 3 and 4, respectively at 200 Oe field strength. The insets show the $M_{irr}$  for 200 Oe and 2000 Oe field strength. (e) $T_{C}$ and $\theta_{CW}$ values for all samples. (f) $\mu_{eff}$ values for all samples. (g) $M_{irr}$ as a function of temperature for data collected at 2000 Oe (inset shows values of $M_{irr}$ at lowest temperature (empty squares) and saturation magnetization values(filled circles)). (h) M vs H curves for samples 2 (at 2 K), 3 (at 5 K) and 4 (at 2 K).}
\label{mag}
\end{figure}

Fig.\ref{mag}(a) shows the dc susceptibility of all the samples collected during the field cooled (FC) and zero field cooled cycles (ZFC) with an applied magnetic field of $2000Oe$. The value of the $T_{C}$ was obtained from the minima in the temperature derivative of the susceptibility data of the ZFC cycle. It is important to note that the value of the $T_{C}$ as shown for the case of samples 3 and 4, do not change with the applied field, Figs.\ref{mag}(a,c,d).
With decrease in temperature, all the samples show a broad maximum. In the case of sample 1, during ZFC two features are observed, one $\sim 144K$ and a broad one $\sim 25K$. It is well known that this compound shows strong magneto-crystalline anisotropy with easy axis lying in the \textit{ab} plane\cite{joy1998,durairaj2006}. These broad features observed in the dc susceptibility data are due to such anisotropy existing in this compound. This is also well depicted from the large bifurcation in the FC and ZFC data. Further, in the case of samples 3 and 4, we observe a broad maxima with the increment in the field from 200 to 2000 Oe, Figs.\ref{mag}(a,c,d).

In the high temperature region, above $T_{C}$, the Curie-Weiss fit to the temperature dependent inverse susceptibility data is shown in Fig.\ref{mag}(b). The Curie Weiss temperature ($\theta_{CW}$)is found out by extrapolating the intercept on the temperature axis and it is found to be positive. This suggests that the ferromagnetic exchange correlations exist in all the samples. The Curie Weiss temperature is found to decrease as one goes from sample 1 to sample 2. With Cr doping, we observe an increment in the Curie Weiss temperature suggesting the strengthening of the ferromagnetic interactions. The effective magnetic moment obtained from Curie constant is observed to increase as one goes from sample 1 to sample 2 and later on it decreases, Fig\ref{mag}(f).

The FC and the ZFC curves exhibit bifurcation for all the samples. This suggests magnetic anisotropy existing in all the compounds. Magnetic anisotropy arises due to magnetic crystalline anisotropy, short range magnetic correlations and/or spin orbit coupling. In the case of sample 1, reports \cite{cao1997,klein1996} show magnetic anisotropy with easy axis lying in the basal plane. The difference between the field cooled and zero field cooled magnetisation is represented as $M_{irr}$ is shown in Fig.\ref{mag}(g). The $T_{irr}$ values were found from the $M_{irr}$ vs temperature curves. $T_{irr}$ marks the temperature at which the FC and ZFC curves start to bifurcate. It is observed that the values of $T_{irr}$ for samples 1, 2, 3 and 4 are around 155 K, 38 K, 87 K and 96 K, respectively.

In the case of sample 1, the $M_{irr}$ increases rapidly below $T_{irr}$ upto $\sim$110K, Fig\ref{mag}(g). The increment is smaller in the temperature range 110K-18K and below $\sim$18K, there is sharp increase. From the behaviour of $M_{irr}$, we observe that the magnetic anisotropy decreases as one goes from sample 1 to sample 2. With Cr doping, i.e. as one goes from sample 3 to sample 4, the increase in the $M_{irr}$ suggests an increment in the magnetic anisotropy. It is found that the magnetic anisotropy for sample 3 and sample 4 is more than that of sample 1 at the lowest collected temperature, as compared to the behaviour above 26 and 32 K, respectively. In all the samples, with decrease in temperature, we observe a continuous increment in the FC susceptibility, Fig\ref{mag}(a). This suggests large coercivity at low temperature. From this behaviour, coercivity at the lowest collected temperature can be written in the following order {\ce{Sr_{0.5}Ca_{0.5}RuO_{3}} $<$ \ce{SrRuO_{3}} $<$ \ce{Sr_{0.5}Ca_{0.5}Ru_{0.95}Cr_{0.05}O_{3}} $<$ \ce{Sr_{0.5}Ca_{0.5}Ru_{0.9}Cr_{0.1}O_{3}}}. This behaviour can be further substantiated for samples 2, 3 and 4 from the M vs H behaviour collected at 2K, Fig.\ref{mag}(h). The behaviour of the shapes of the FC and ZFC susceptibility curves and the M vs H curves suggest that magnetic anisotropy plays significant role in these compounds. The effect of magnetic anisotropy is further evident from the decrement in the $M_{irr}$ with increase in the applied field as seen in the case of samples $3\& 4$, Fig.\ref{mag}(insets of c and d).

The isothermal magnetization data for samples 2 to 4 was collected at 2K. All the samples show hysteresis,Fig.\ref{mag}(h) suggesting the samples to be in ferromagnetic state. The magnetisation does not saturate even at 5 T. The value of saturation magnetization($M_{s}$) was obtained by plotting M vs 1/H and extrapolating the x-axis to zero. The values of $M_{s}$ for different samples are shown in inset of Fig.\ref{mag}(g). For the case of sample 1, the value of $M_{s}$ was obtained from the literature\cite{nithya2009}. It is observed that the value of $M_{s}$ increases as one goes from sample 2 to sample 4.

As shown in Fig.\ref{mag}(e), the value of $\theta_{CW}$ and $T_{C}$ are same for sample 1 but different for other samples. Also, the inverse susceptibility curves for samples 2, 3 and 4 show a downturn above the ferromagnetic ordering temperature, Fig.\ref{mag}(b) but a straight line is seen in case of sample 1. These observations hint at the possibility of local magnetic ordering in the doped samples. In samples 2 to 4, the Ca and Cr dopant ions are randomly substituted inside the polycrystalline sample which can create regions of local magnetic ordering in these systems above the respective Curie points. To investigate the origin of this behaviour we fitted the inverse susceptibility data with the modified Curie-Weiss equation\cite{dayal2014evolution} given as

\begin{equation}
\chi^{-1}\propto(T-T_{R})^{1-\lambda}
\label{eq2}
\end{equation}

where $\lambda$ is the Griffiths exponent and $T_{R}$ is the temperature at which the system shows transition from Griffiths phase to the ferromagnetic phase\cite{neto1998,rathi2018}. The exponent $\lambda$ takes a value $0<\lambda\leqslant1$ in the Griffiths phase region and is equal to zero in the paramagnetic phase. The best estimate of $T_{R}$ is the Curie temperature $\Theta_{CW}$ as the equation(\ref{eq2}) gives $\lambda=0$ in the paramagnetic phase for $T_{R}=\Theta_{CW}$\cite{kumar2020}.

\begin{figure}[h!]
 \vspace{1ex}
\includegraphics [width=0.95\linewidth]{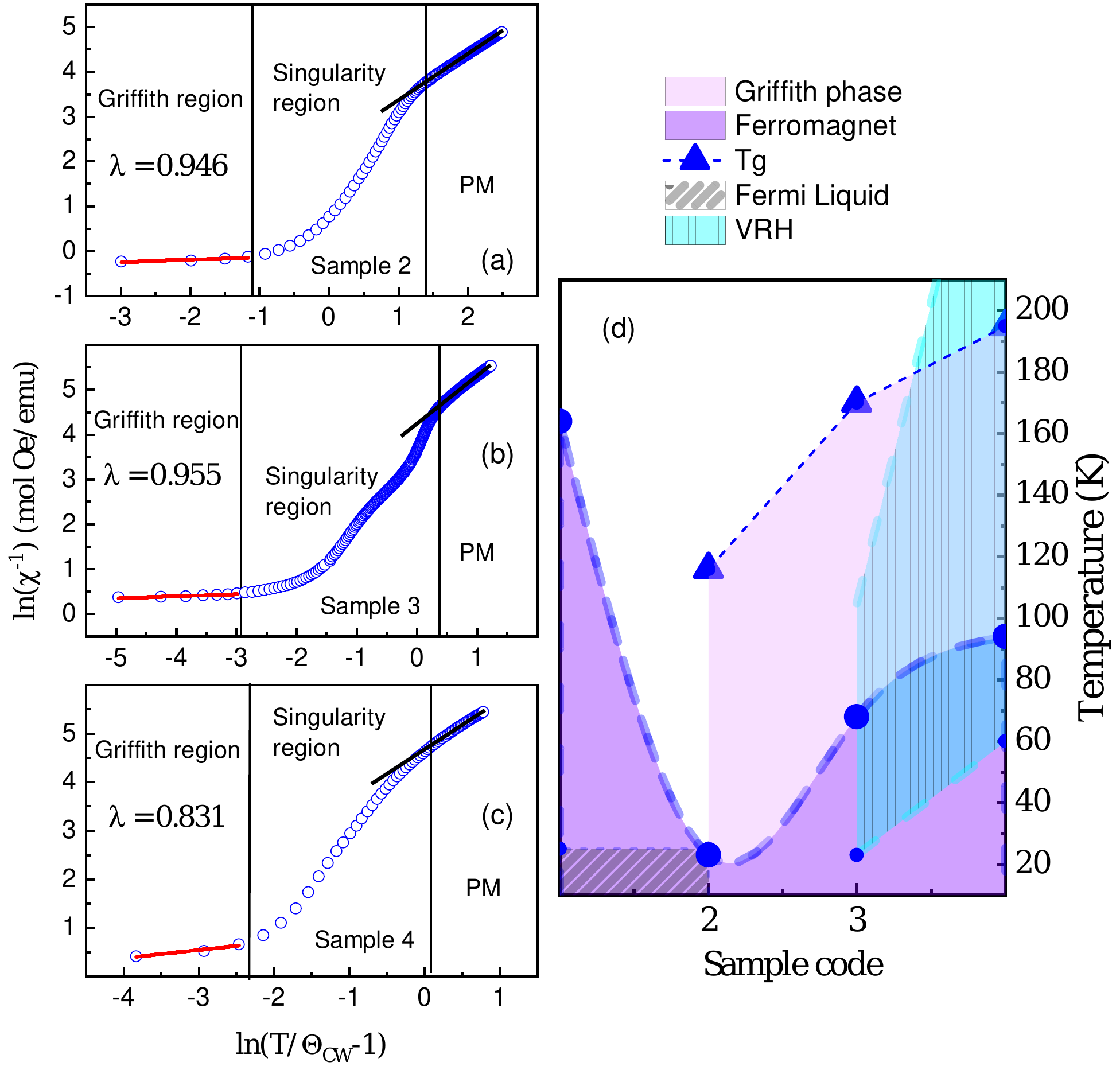}
\vspace{-1ex}
\caption{(a,b,c) Natural log of $\chi^{-1}$ is plotted against the natural log of ($T/\Theta_{CW}-1$) for the doped samples. The value of the Griffiths exponent obtained from fitting(solid red lines) is mentioned.(d) Phase diagram showing regions of magnetic and transport properties.}
\label{gp}
\vspace{-1ex}
\end{figure}

Fig.\ref{gp} shows the log of inverse susceptibility vs the log of ($T/\Theta_{CW}-1$) for samples 2, 3 and 4. The temperature region showing linear behaviour(marked as Griffiths phase in figure) is fitted with straight line to extract the Griffiths exponent. $\lambda$ takes the values 0.946, 0.955 and 0.831 for samples 2, 3 and 4, respectively. The positive values of $\lambda$ confirm the presence of Griffiths phase in the doped samples. $T_{G}$ is the temperature below which the local ordering starts in the sample. Values of $T_{G}$ are estimated as the points where $1/\chi$ starts deviating from the straight line behaviour seen in the high temperature region(paramagnetic regime), as shown in Fig.\ref{gp}. Based on the values of $T_{G}$ and $\theta_{CW}$, a magnetic phase diagram is made showing the extent of the Griffiths phase region for the doped samples, Fig.\ref{gp}(d). As one goes from sample 2 to 4, the value of $T_{G}$ is found to increase from 116 K to 195 K. The width of the temperature region of Griffiths phase is $\sim$ 93 K for sample 2. It increases for samples 3 and 4 to $\sim$ 102 K and $\sim$ 101 K, respectively. The similar temperature range shows that the extent of local magnetic ordering is similar in samples 3 and 4.

When random disorder is created in a ferromagnetic material, regions of local magnetic ordering termed as rare regions\cite{ubaid2010} exist inside the paramagnetic bulk above the Curie temperature. Then the total susceptibility of the system is the sum of the susceptibilities due to the rare regions and the paramagnetic bulk. If the local regions have ferromagnetic ordering, the total susceptibility of the sample will have a higher value as compared to the paramagnetic susceptibility. As a result, a downturn is seen in the $1/\chi~vs~T$ curve at the temperature where the rare regions start to order.

The strength of the local magnetic ordering can be estimated from the difference in the T$_{C}$ and $\Theta_{CW}$ values. Sample 1 has no substitutional disorder or local ordering and shows the same value of T$_{C}$ and $\Theta_{CW}$ whereas a difference of $15 K$ is observed for sample 2. Doping of Cr does not alter this difference significantly. The two values differ by $13 K$ in case of both samples 3 and 4. This shows that the emergence of local magnetic ordering in the system is mainly due to the A site substitution.

\subsection{Transport studies}

\begin{figure}[h!]

\includegraphics[width=0.98\linewidth]{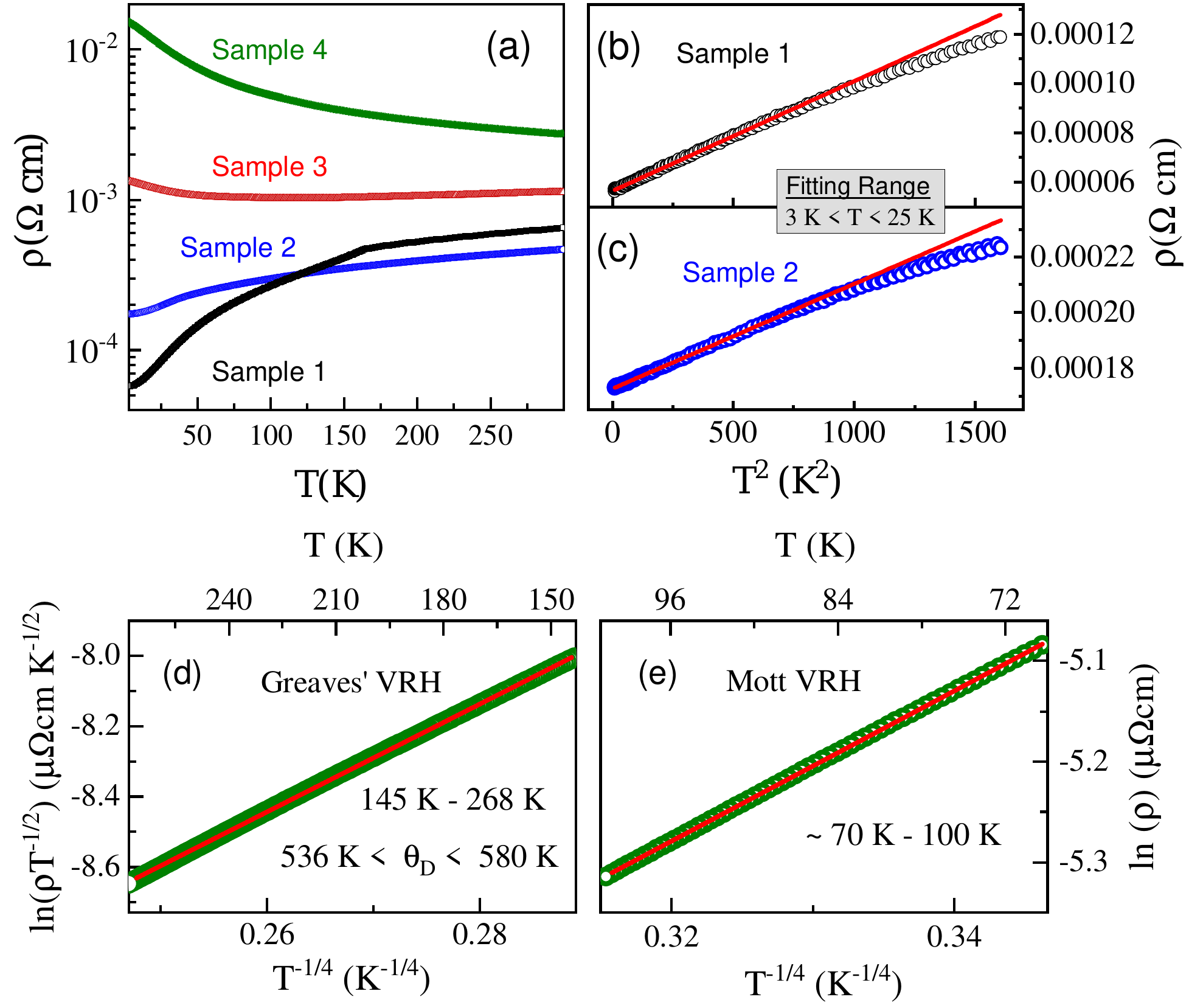}
\vspace{-1ex}
\caption{(a)  Resistivity of samples 1 to 4 as a function of temperature. (b) and (c) Low temperature resistivity of samples 1 and 2 respectively. Solid lines show the $T^{2}$ dependence (eq.\ref{FLT}). (d) and (e) show the log of $\rho T^{-1/2}$ and log of $\rho$ respectively, plotted against $T^{-1/4}$ showing different VRH mechanism. Solid lines are straight line fits.}
\label{resistivity1}
\vspace{-1ex}
\end{figure}

Fig.\ref{resistivity1}(a) shows the temperature dependent resistivity for samples 1 to 4. At room temperature the resistivity values go in the order, sample 2 $<$ sample 1 $<$ sample 3 $<$ sample 4. Metal to non metallic transition is observed as one goes from sample 1 to sample 4. At the lowest collected temperature the behaviour of resistivity of samples 1 and 2 reverses and as we go from sample 2 to 4, it becomes non-metallic. In the temperature region of study, samples 1 and 2 show metallic behaviour and there is crossover in the resistivity values of samples 1 and 2 below $\sim$ 120 K. A kink is observed $\sim$ T$_{c}$ for samples 1 and 2. Additionally, a slope change is observed $\sim$ 70 K in case of sample 1. In the case of sample 3, a metal to non-metallic transition is observed $\sim$ 116 K with decrease in temperature. Sample 4 shows non-metallic behaviour in the entire temperature range of study. We now look closely into the resistivity behaviour of each and every sample. Considering the fact that the samples 3 ad 4 lie in the transition region, one can expect the non metallic behaviour more of disorder induced rather than Arrhenius type of behaviour and the disorder being induced due to non metallic regions present in the metallic matrix.

In literature\cite{allen1996}, the kink observed in resistivity around $T_{C}$ in case of the sample 1 is attributed to the scattering associated with the short range spin fluctuations around $T_{C}$. This is as per Fisher-Langer theory\cite{cao2005}. Reports show that in the case of this compound, the resistivity increases upto 900 K, violating the Mott-Ioffe-Regel limit\cite{emery1995}. The behaviour of the resistivity data was understood using different models. In case of samples 1 and 2, for T $<$ 25 K, the resistivity data follows the Fermi liquid behaviour with the functional form
\begin{equation}
\rho = \rho_{0} + AT^{2}
\label{FLT}
\end{equation}
where $\rho_{0}$ is the residual resistivity and A is the temperature coefficient. The values of coefficient A obtained from fitting are $4.43 \ast10^{-8}\Omega$cm and $3.77\ast10^{-8}\Omega$cm, respectively, Fig.\ref{resistivity1}(b and c). These values are close to the values obtained in the case of strongly correlated systems\cite{cao2005}.

Two different conduction mechanisms were found in case of sample 4 in two temperature regions. In the intermediate temperature region, the resistivity data is fit with Greaves' Variable Range Hopping (VRH) mechanism\cite{greaves1973}, following the equation \ref{greaves}

\begin{equation}
\rho (T) = AT^{1/2}~exp(T_{G}/T^{1/4})
\label{greaves}
\end{equation}

where $T_{G}$ is the characteristic temperature of Greaves' VRH and A is the pre-exponential factor. Mott suggested that disordered systems have random potentials in the bulk, because of which the electrons show variable range hopping($ln\rho \sim T^{-1/4}$) in the lower temperature regimes\cite{mott1985,mott2012}. At higher temperature range, the thermal energy is sufficient to hop electrons between the nearest neighbour sites irrespective of the potentials and the system shows activated conduction ($ln\rho \sim T$). For the intermediate temperature range, where both optical and acoustic phonons are responsible for the hopping of electrons, Greaves' VRH conduction mechanism is expected. The temperature range of the Greaves' VRH mechanism is typically $\theta_{D}/4 \leq T \leq \theta_{D}/2$ where $\theta_{D}$ is the Debye temperature of the sample\cite{greaves1973,kumar2007}. The straight line fit between $ln(\rho T^{-1/2})$ and $T^{-1/4}$ (Fig.\ref{resistivity1}d) is found in the temperature range 145 K - 268 K, giving the value of $\theta_{D}$ in the range 536 K to 580 K.

In the lower temperature range, $70 K - 100 K$, the resistivity data is found to exhibit Mott VRH conduction mechanism following equation \ref{vrh}. Natural log of $\rho$ shows straight line behaviour when plotted against $T^{-1/4}$, fig\ref{resistivity1}(e).

\begin{equation}
\rho (T) = \rho_{0}~exp(T_{0}/T)^{1/4}
\label{vrh}
\end{equation}

It is interesting to note that the upper limit of Mott VRH conduction mechanism is same as $T_{C}$ for sample 4. Also, the unit cell volume shows a change of slope near 100 K(discussed in next section). The value of $T_{0}$ is found to be 3747 K. These results show that disorder is playing important role in case of sample 4. It has been shown in literature that the characteristic temperature, $T_{0}$ is inversely related to the density of states at Fermi level. To understand the origin of these transport properties, photo-emission spectroscopy experiments will be helpful.

\subsection{Structural Studies}

\begin{figure}[h!]
 \vspace{-5ex}
\includegraphics[width=0.98\linewidth]{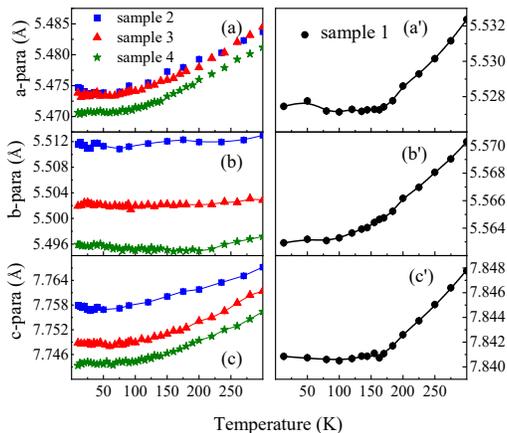}
 \vspace{-34ex}
\caption{Lattice parameter values of all the samples as a function of temperature. (a'), (b') and (c') show the parameters for sample 1 and (a), (b) and (c) show that for samples 2, 3 and 4.}
\label{latpara}
 \vspace{-1ex}
\end{figure}

\begin{figure}[h!]
\vspace{-5ex}
\includegraphics[width=0.98\linewidth]{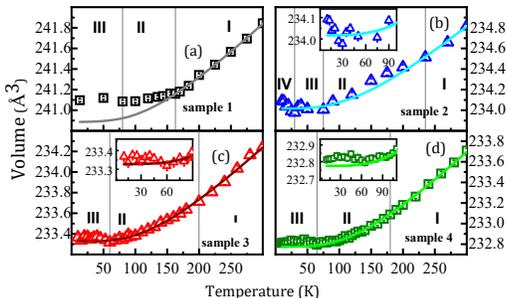}
\vspace{-47ex}
\caption{Variation of unit cell volume with temperature for samples 1, 2, 3 and 4 is plotted in (a), (b), (c) and (d), respectively. Solid lines show the curves generated by the Debye model.}
\vspace{-1ex}
\label{vol}
\end{figure}

To understand the link between crystal structure, transport properties and magnetism, we now analyse the temperature dependent structural parameters. In Fig.\ref{latpara} we show temperature dependence of the lattice parameters for all the samples. There is significant change in the lattice parameter values with temperature and composition. The temperature dependent volume of the unit cell for all the samples is shown in Fig.\ref{vol}. To understand this behaviour, fitting was carried out using the anharmonic Debye model\cite{kiyama1996,ramirez2012} which gives the lattice contribution to the unit cell volume following the equation

\begin{equation}
\label{eq4}
V=V(T=0) + \dfrac{9\gamma Nk_{B}}{B}T\left(\dfrac{T}{\theta_{D}}\right)^{3} \displaystyle{\int_{0}^{\theta_{D}/T} \dfrac{x^3}{e^x -1} dx}
\end{equation}

where $V(T=0)$, $\theta_{D}$, $\gamma$ and $B$ are the volume at $0K$, the Debye temperature, the Gr\"{u}neisen parameter and bulk modulus, respectively. The fitting was done in the temperature range extending from $T_{C}$ to 300 K for all the samples under study. The results of the fit are given in the Table \ref{table:debye}. For sample 1, the values of $\theta_{D}$ and $9\gamma Nk_{B}/B$ obtained from fitting are in line with the earlier reports\cite{kiyama1996}. As one goes from sample 1 to sample 4 the increment in the $\theta_{D}$ suggest increment in the temperature required for phonon excitation which further reflects increment in the strength of the bonds. The value of $\theta_{D}$ obtained from fitting of the resistivity data (fig.\ref{resistivity1}d) is in line with the value obtained from fitting of volume data for the case of sample 4.

\begin{table}[h!]
\caption{Parameters obtained from fitting of equation~\ref{eq4}}
\centering
\setlength{\tabcolsep}{5.4pt}
\renewcommand{\arraystretch}{1.6}
\begin{tabular}{c c c c c c}
\hline\hline
Composition & & $V(T=0K)$ & $\theta_{D}$ & $9\gamma Nk_{B}/B$ & M$_{sat}$ \\
 & & (\AA$^{3}$) & (K) & (\AA$^{3}$/K) & ($\mu_{B}$) \\
\hline
Sample 1 & & 240.89 & 522.29 & 0.01932 & 1.52$^{*}$ \\
Sample 2 & & 234.02 & 542.85 & 0.01661 & 0.50 \\
Sample 3 & & 233.33 & 555.03 & 0.01935 & 0.67 \\
Sample 4 & & 232.77 & 558.35 & 0.01968 & 0.75 \\
\hline
\end{tabular}
$^{*}$The value is taken from reference \cite{nithya2009}
\label{table:debye}
\end{table}

\begin{figure}[h!]
\vspace{-3ex}
 \includegraphics[width=0.95\linewidth]{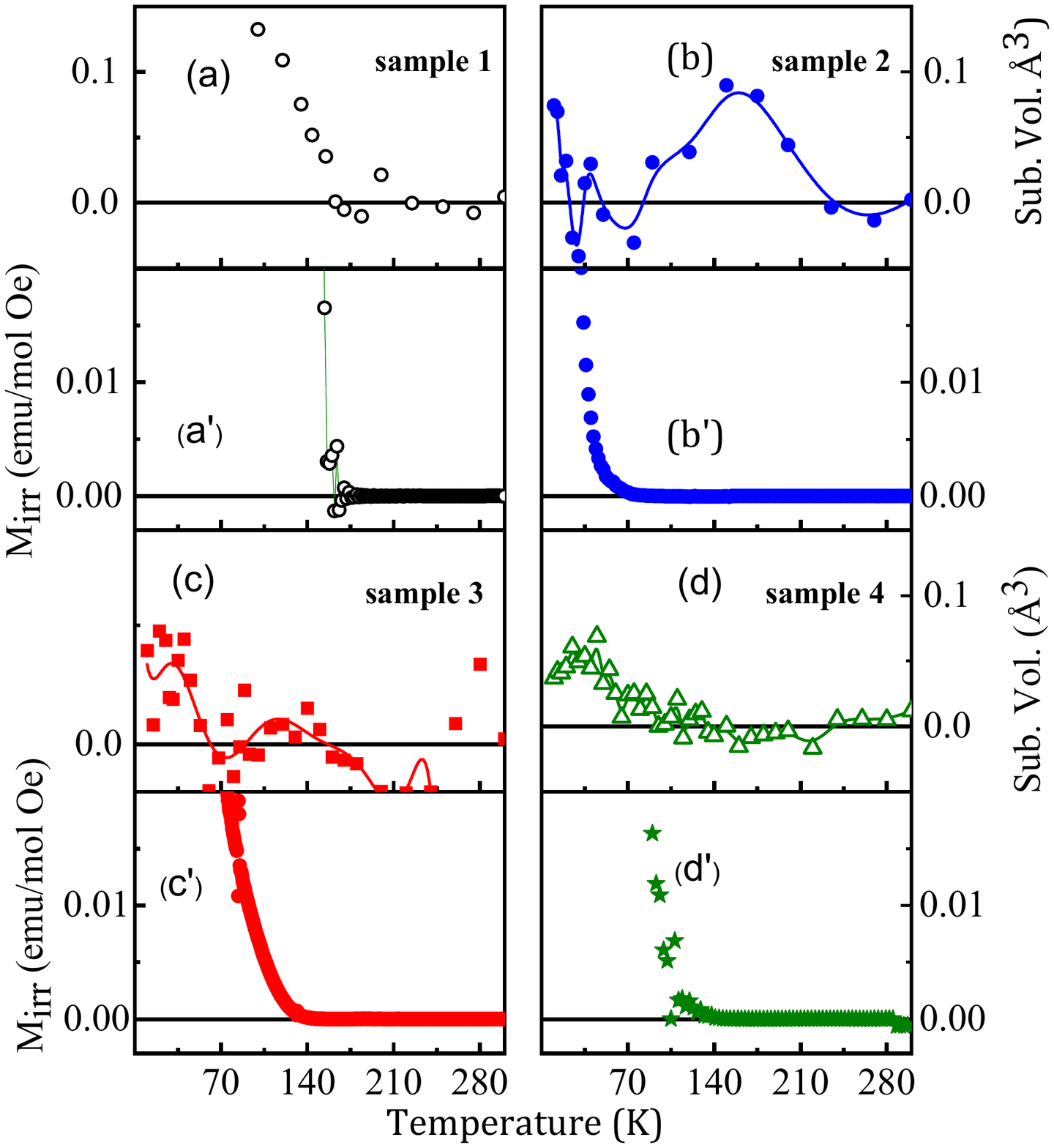}
 \vspace{-20ex}
 \caption{(a), (b), (c) and (d) show the difference of the experimental unit cell volume values and the fit using Debye model for samples 1, 2, 3 and 4 respectively. (a'), (b'), (c') and (d') show the $M_{irr}$ vs temperature for samples 1, 2, 3 and 4, respectively.}
\label{subvol}
\vspace{2ex}
 \end{figure}

The Debye curve, Fig.\ref{vol} generated in the entire temperature range was subtracted from the experimental data. Based on the temperature dependent behaviour of volume, different regions can be set for each samples. We now present the results of the structural parameters for all the samples.

From the Fig.\ref{subvol}(a-d), it is clear that the subtracted volume shows an increase in all the compounds around Tc. This is in line with the rise in the Mirr vs Temperature curve Fig.\ref{subvol}(a'-d’). The sample 2 shows a hump around 150 K. These behaviours suggest connection between ferromagnetism, magnetic anisotropy and the structural parameters.

\begin{figure}[h!]
 \vspace{-8ex}
\includegraphics[width=0.98\linewidth]{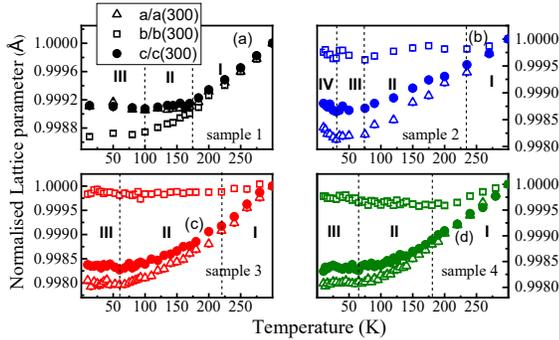}
 \vspace{-42ex}
\caption{(a), (b), (c) and (d) show the lattice parameter values normalized at 300K for samples 1, 2, 3 and 4, respectively.}
\label{normlatpara}
 \vspace{-1ex}
\end{figure}

In Fig.\ref{normlatpara}(a-d), the lattice parameters have been normalised with respect to RT. For the case of sample 1, it is interesting to note that a and c-parameters change by the same amount in all the three regions, Fig.\ref{normlatpara}(a). In region I, the change in the b-parameter is slightly more as compared to the other two parameters and it deviates significantly in regions II and III. A weak hump is also observed around 50 K in the structural parameters. This is in line with other studies\cite{kiyama1996}.

\begin{figure}[h!]
\vspace{2ex}
\includegraphics[width=0.95\linewidth]{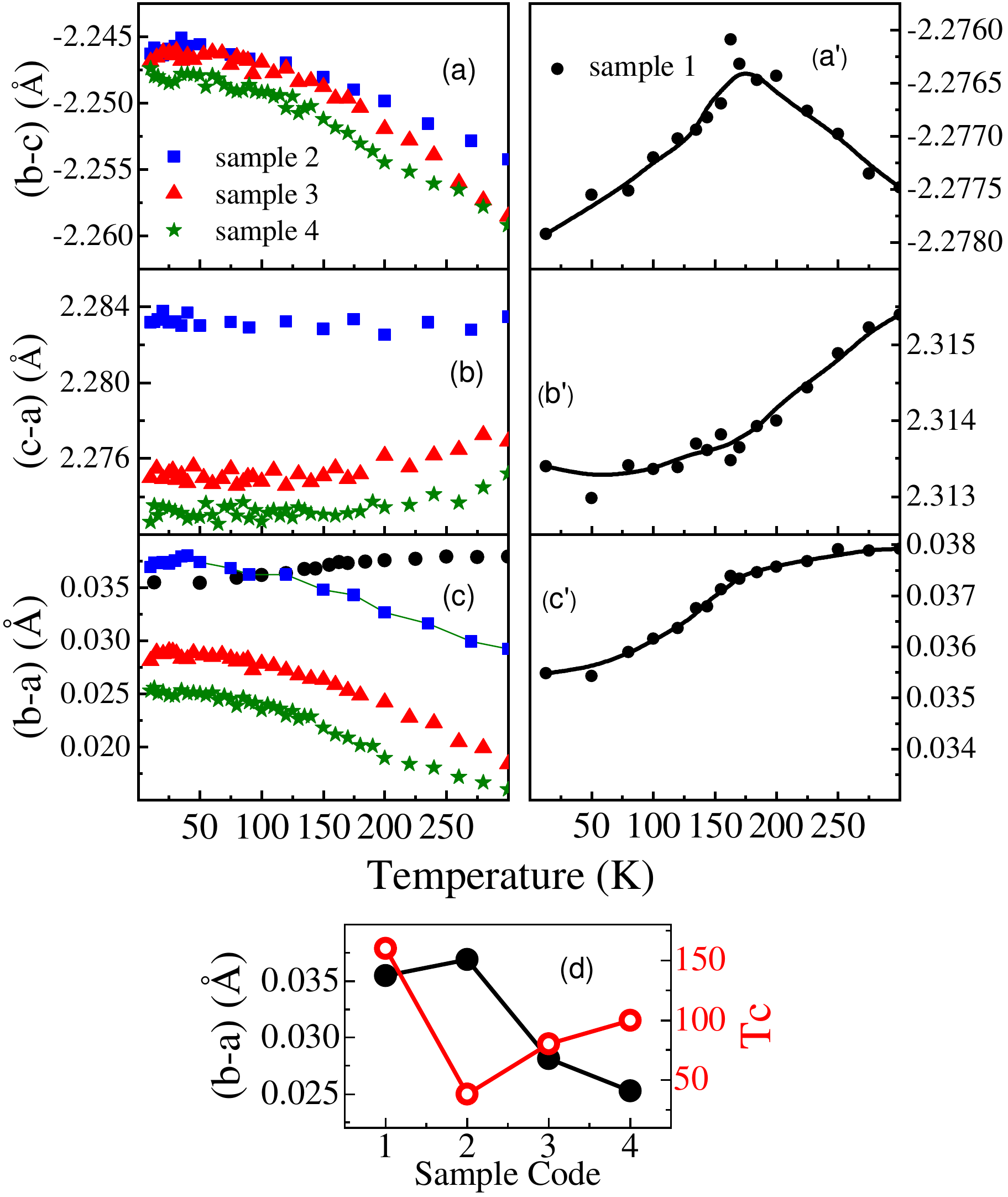}
\vspace{-1ex}
\caption{Temperature dependent difference in the lattice parameter values . (a), (b) and (c) show (b-c), (c-a) and (b-a), respectively for samples 2, 3 and 4. (a'), (b') and (c') shows data for sample 1. (d) shows the variation of orthorhombic strain at lowest temperature collected for all samples(filled circles) along with $T_{C}$(open circles).}
\label{diff2}
\vspace{-2ex}
\end{figure}

Unlike sample 1, the nature of the change and the percentage change in the lattice parameters in sample 2 is different, Fig.\ref{normlatpara}(b). In this compound, in region I, all the parameters decrease linearly. In region II, the b and c-parameters show a hump around 150 K. The a and c parameters deviate from each other and change in the b-parameter is significantly different from other two parameters. From this it is clear that the change in the a and c parameters is markedly increased as compared to the b parameter. In all the three lattice parameters, a small peak is observed around Tc and in region IV, all the parameters show an increment. Further, it is interesting to note that in the (b-a) vs temperature and (b-c) vs temperature plots, Fig.\ref{diff2}, a small peak is observed around $T_{C}$ and a kink around 125 K. The kink observed around 125 K in sample 2 is in line with the slope change observed in the resistivity data around this temperature, Fig. \ref{resistivity1}(a).

In Fig.\ref{diff2}, when we compare graph obtained for samples 1 and 2, we observe that (b-c) and (c-a) values for the sample 1 are larger than that of sample 2. While in the case of temperature dependent (b-a) plot, around 125 K, the (b-a) value of sample 1 is lower than that of sample 2. It is interesting to note that the resistivity of sample 1 also shows a similar crossover with the resistivity of sample 2 around 125 K, Fig.\ref{resistivity1}(a). This shows that orthorhombic strain governs the transport behaviour in a wide temperature range, even in the region where the resisitvity is linear.

We now look into the behaviour of the normalised lattice parameters as a function of temperature for the case of sample 3. It is interesting to note that as one goes from sample 2 to sample 3, the width of region I extents from 65 K to 80 K, Fig.\ref{normlatpara}. The hump in the b-parameter that is observed in sample 2 has now reduced in sample 3 and as the sample enters region III, the b parameter shows a small peak around 30 K. The resistivity data also shows an increment in region III (below 60 K) and metal to non metallic transition below 60 K as one goes from sample 1 to sample 3 sequentially (Fig.\ref{resistivity1}(a)). It is interesting to note that the value of the resistivity in the case of sample 3 lies in the resistivity range shown by semi metals. However, low temperature high resolution photoemission and angle resolved photoemission on the single crystalline compound will be helpful to unravel the nature of this material. From these behaviours, it appears that b-parameter plays significant role in governing the resistivity. In the (b-c) vs temperature curve, below 150 K, the samples 2 and 3 values remain almost the same, Fig.\ref{diff2}. There is a significant difference in the temperature dependent (c-a) and (b-a) behaviours for samples 2 and 3.

In Fig.\ref{normlatpara}(d), we show the temperature variation of the normalised lattice parameters of sample 4. All the three lattice parameters are found to decrease in region I. In comparison to samples 2 and 3, the extent of region I has increased to 180 K. In region II, the lattice parameters a and c show a decrement while the b-parameter shows an increment. This is also accompanied by an increment in the resistivity data, Fig.\ref{resistivity1}(a). As the sample enters region III, around 40 K, a peak in the b-parameter(Fig.\ref{normlatpara}(d)) and an increment in the resistivity are observed. The difference in the lattice parameters namely, (b-c), (c-a) and (b-a) shows a decrement in its values in comparison to the rest of the samples, Fig.\ref{diff2}.

In Fig.\ref{diff2}(d) we observe that the behaviour of the Tc and the orthorhombic strain obtained at the lowest collected temperature as a function of composition shows opposite behaviours. Hence depicting the direct link between the orthorhombic strain and Tc. Detailed band structure calculations will be helpful in understanding the occupancies of the electrons and deciphering the interactions that govern the properties of these compounds. This also could provide clue on the behaviour of Tc and the origin of ferromagnetism with doping.

To understand the origin of the behaviour of the structural parameters, it is important to understand the valence state of Ru and Cr ions. In order to find out the valence state of Ru and Cr ions, the details of the ionic radii are studied. The ionic radii of \ce{Ru^{4+}} is $0.62\AA$, \ce{Ru^{5+}} is $0.565\AA$, \ce{Cr^{3+}} is $0.615\AA$ and \ce{Cr^{4+}} is $0.55\AA$\cite{dabrowski2005,shannon1969}. Hence, if there is \ce{Ru^{5+}} and \ce{Cr^{3+}} existing in this compound, the lattice parameter is expected to increase based on the ionic model. But opposing trend is observed experimentally, Fig.\ref{latpara}. Hence, the valence state of Ru is expected to be 4+ in all the samples and in samples 3 and 4, the valence state of Cr is 4+.

To investigate further into this, we look into the results of the magnetic studies. In case of \ce{SrRuO_{3}}, the electronic configuration of $Ru^{4+}$ is $4d^{4}$ (with 2 unpaired electrons). Hence the magnetic moment ($\sqrt{n(n+2)}$) expected will be $2.84\mu B$. The experimental magnetic moment value is similar to that of $Ru^{4+}$. Hence it is expected that Ru is in 4+ valence state in \ce{SrRuO_{3}}. In the case of rest of the samples, had there been $Ru^{5+}$ and $Cr^{3+}$, the effective magnetic moment is expected to be more than that obtained for $Ru^{4+}$ and $Cr^{4+}$ ions but experimentally, opposite was observed. Hence, these results are in accordance with Ru and Cr being in 4+ valence state. However, to find the valence state in these compounds xps measurements were carried out for samples 1 and 4.

\subsection{Electronic structure studies}

\begin{figure}[h!]
\vspace{-1ex}
\includegraphics[width=0.9\linewidth]{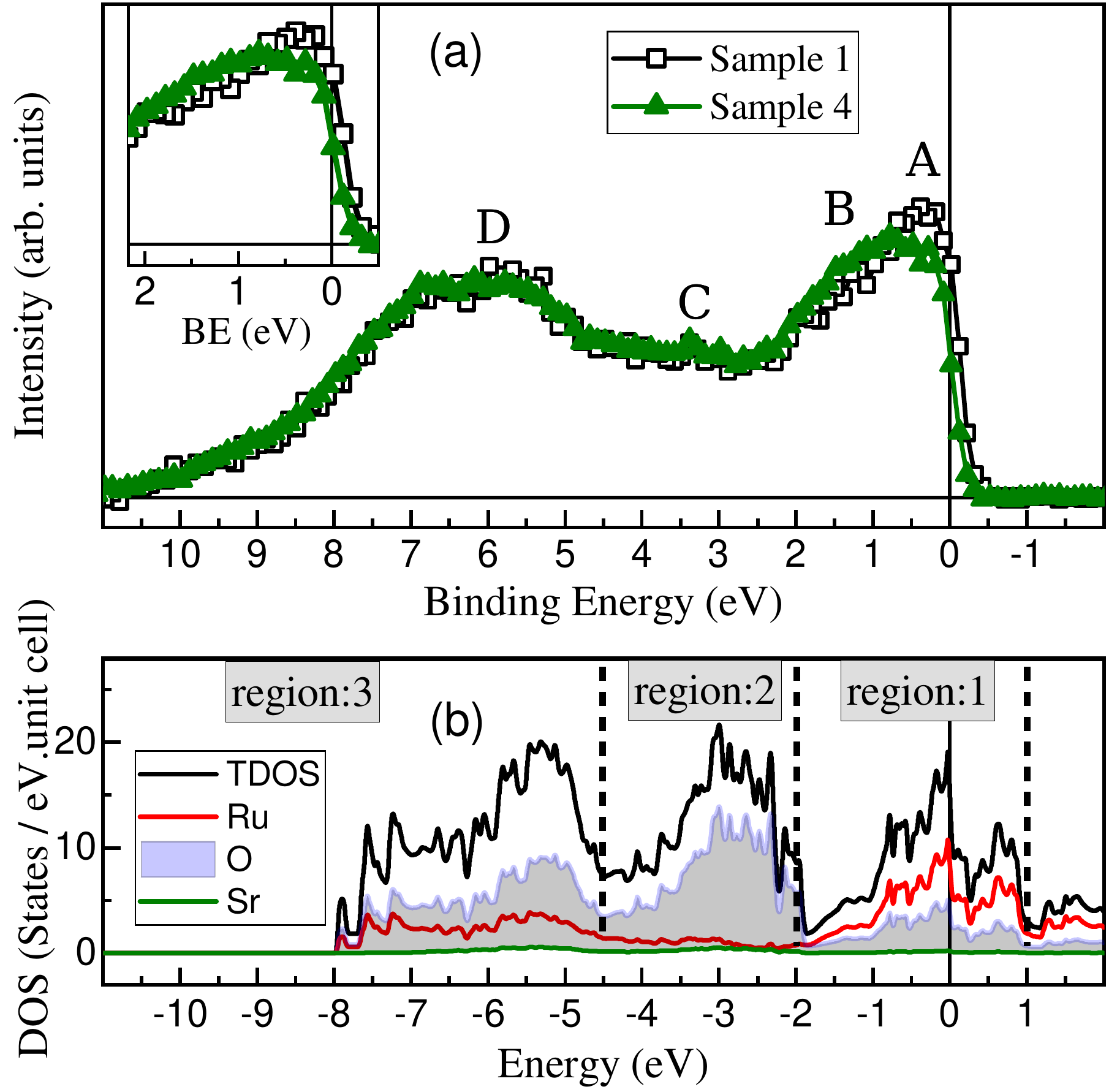}
\vspace{-1ex}
\caption{(a) Room temperature xps valence band spectra of samples 1 and 4. (b) Total and partial density of states (DOS) calculations of SrRuO$_{3}$ done using DFT method.}
\label{vb}
\vspace{-1ex}
\end{figure}

In Fig.\ref{vb}(a), we show the room temperature xps valence band spectra collected for samples 1 and 4. It is important to note that the xps represents the bulk electronic structure. The features in the valence band spectra are labelled as A,B, C and D. As we go from sample 1 to sample 4, the intensity at the Fermi level and feature A is found to decrease and an increase in the intensity of feature B is observed.

To identify the features in the valence band spectra band structure calculations were carried out on \ce{SrRuO_{3}} compound under DFT, using the structural parameters obtained experimentally. Our calculation results show that the total density of states (TDOS) can be divided into 3 regions. Regions 1, 2 and 3 mark the energy range 1 to -2 eV, -2 to -4.5 eV and below -4.5 eV, respectively. In region 1, there is significant contribution from Ru 4d states and less contribution from O 2p partial DOS. The region 2 is dominated by O 2p states and there is no significant contribution from Ru 4d states. The region 3 is dominated by O 2p states and there is a weak contribution from Ru 4d states. The contribution of Sr states is found to be insignificant in these regions.

In sample 1, the features A and B that cover the region 1, can be attributed to dominant contribution from Ru 4d states in comparison to O 2p states. We know that for Al K$\alpha$ source, the photoionization cross section for Cr 3d states is one order of magnitude smaller than that for the Ru 4d states and also the number of electrons in the d states are expected to decrease with Cr doping.  In such situation one expects decrement in the overall intensity in the region where d-states contribute. But experimentally, we observe, a reduction in the intensity at the Fermi level and feature A and an increment in the intensity of feature B. Such behaviour suggests spectral weight transfer of the intensity from feature A and at the Fermi level to feature B. With Ca and Cr doping, disorder is expected to be induced that affect the network of Ru-O-Ru bonds. Further the radial extent of Cr 3d states is expected to lead to the reduction in the band width and increment in the electron electron correlation strength. So the spectral weight transfer of the intensity from feature A to B and a small shift in these features towards higher binding energy suggest the decrement in the delocalization of the electrons in the d states and increment in its localization due to Cr doping and hence the states that contribute close to the Fermi level are expected to shift to higher binding energy. The decrement in the number of electrons due to the doping effects could also lead to the decrement in the intensity at the Fermi level. Such behaviour has also been observed in literature and the features A and B have been attributed to coherent and incoherent features, respectively\cite{singh2007}. The coherent feature corresponds to the delocalized states and the incoherent feature to the localization due to electron electron correlation effects. On connecting these results with the transport properties of the samples 1 and 4, we understand that the metal to non metal transition observed with increase in doping is associated with disorder, electron-electron correlation and reduction in the number of electrons at the Fermi level.

\begin{figure}[h!]
\vspace{-5ex}
\includegraphics[width=0.95\linewidth]{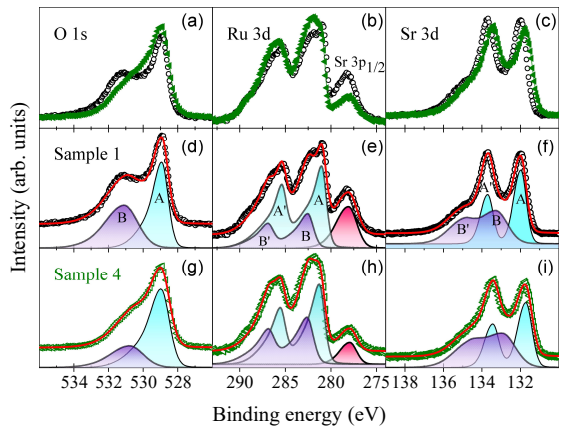}
\vspace{-30ex}
\caption{(a), (b) and (c) : Room temperature O 1s, Ru 3d and Sr 3d core level spectra, respectively for samples 1 and 4. (d) and (g) : The fitting of O 1s spectra for samples 1 and 4, respectively. (e) and (h) : The fitting of Ru 3d spectra for samples 1 and 4, respectively. (f) and (i) The fitting of Sr 3d spectra for samples 1 and 4, respectively.}
\label{corelevels}
\vspace{-1ex}
\end{figure}

To understand the transport behaviour from the core level spectra and also the valence state of the Ru ions, core level studies were carried out. All the core level spectra were normalised to the area under the curve after background subtraction.

Fig.\ref{corelevels} shows the room temperature core level spectra collected for samples 1 and 4. In the case of the O 1s spectra, both the samples display two features labelled as peak A and B around 529 and 531 eV, respectively. The feature A is intense as compared to the feature B. In most of the perovskite compounds, the feature A has been attributed to the signal from the oxygen ions intrinsic to the sample while the weak feature B has been attributed to the signal from the surface oxygen and/or adsorbed impurities present on the sample. In the present compounds, the surface cleaning of the samples were done by scraping until the intensity of feature B is minimum with respect to feature A. We observe that the intensity ratio of feature A to feature B is less as compared to that observed in other transition metal oxides\cite{bindu2010,bharath2019,maurya2017}. This suggest the signals from feature B to be intrinsic to the sample. Similar behaviour has been reported in literature\cite{singh2006,singh2007} and has been attributed to the signals arising primarily due to lattice effects. The different oxygen sites namely apical and the basal oxygen ions in the RuO$_{6}$ octahedra give rise to different Madelung potentials. Considering the fact that the features A and B arise due to lattice effects, we now attribute the features A and B to the signal arising from O1 and O2 sites, respectively. As one goes from sample 1 to 4, the decrement in the separation between the features A and B suggest the decrement in the separation between the average Ru/Cr-O(1) and Ru/Cr-O(2) bonds lengths, thereby leading to the decrement in Madelung potential between both the oxygen sites. In the case of sample 1, it is 0.0146${\AA}$ and for the case of sample 4, it is 0.0059${\AA}$. To understand the behaviour of the shift in the peak position as one goes from sample 1 to sample 4 and also the valence state of Ru ions, we now look into the results of the Ru 3d core level spectra.

In Fig.\ref{corelevels}(e), we show the Ru 3d core level spectra. To understand the behaviour of this Ru 3d spectra each of the spin orbit split peak was fitted with a set of 2 asymmetric peaks labelled as A and B (for Ru 3d$_{5/2}$) and A’ and B’ (for Ru 3d$_{3/2}$). The features A and B correspond to well screened and poorly screened final states, respectively. The well screened feature arises due to the screening of the Ru 3d core hole by the electrons of the ligand through charge transfer. The poorly screened feature arises when there is no such screening of the core hole. Our results show that the ratio of the area under the curve of peak A to peak B decreases as one goes from sample 1 to sample 4. It is also observed that the spectrum of sample 4 is shifted to higher binding energy as compared to sample 1. The behaviour of the ratio of the area under the curve of  peaks A to B suggests that as the sample enters the non metallic state, the screening of the core hole due to charge transfer is reduced thus leading to the increment in the area under the curve of the peak B as compared to peak A. The opposite behaviour observed in the shift in the peak position of O1s and Ru 3d states suggests that as the sample enters the insulating state, there is reduced transfer of electrons from O 2p to Ru 3d states. These behaviours are in line with the structural results where we have observed decrement in the av.Ru-O bond lengths and av. Ru-O-Ru bond angle from sample 1 to sample 4. Considering the fact that the fit matches the experimental spectra in both the samples, we attribute the valence state of Ru to be in 4+ state for both the samples.

The comparison of the Sr 3d core levels of samples 1 and 4 is shown in Fig.\ref{corelevels}(c). Our results show that the intensity of the Sr 3d core level in the case of sample 4 is less as compared to sample 1. This is because of the reduced percentage of Sr in the compound due to Ca doping. To simulate the experimental spectra, for each spin orbit split peaks, two set of peaks were generated that constitute the well screened and poorly screened final states\textbf{\cite{han2002,singh2007}}. The features A and B correspond to the well screened and poorly screened channels, respectively of Sr 3d$_{5/2}$ spin orbit split peak. Similarly, A’ and B’ correspond to that for Sr 3d$_{3/2}$ peak. As one goes from sample 1 to sample 4, the area under the curve of peak A to B is found to reduce. Such reduction suggests the reduction in the hybridization of the Sr 3d and O 2p states due to the distortion in lattice introduced due to doping. We also observe that the av. Sr-O bond length decreases as one goes from sample 1 to 4. Hence it is expected that the peak position of the Sr 3d core level in the case of sample 4 to be shifted towards higher binding energy while in our case opposite behaviour is observed. So, the observed shift towards lower binding energy can be explained due to chemical potential shift towards the valence band. This is expected when there is effective decrement in the number of electrons due to Cr doping.

\section{Summary}

We investigate the structural, transport, magnetic and electronic properties of SrRuO$_{3}$, Sr$_{0.5}$Ca$_{0.5}$Ru$_{1-x}$Cr$_{x}$O$_{3}$ ($\texttt{x}$ = 0, 0.05 and 0.1) labelled as sample 1 to 4. All the compounds (a) stabilize in orthorhombic structure with $\emph{Pbnm}$ space group, (b) exhibit transition from paramagnetic to ferromagnetic state and (c) the magnetization does not saturate even at 5T. Our combined core level and magnetic studies suggest that Ru and Cr are in 4+ valence state and the chemical potential is found to shift towards the valence band due to Cr doping.  At RT, as one goes from sample 1 to sample 4, metal to non metal behaviour is observed. The samples 3 and 4 lies in the transition region from metal to non metallic. At low temperatures,the sample 4 obeys Mott VRH type conduction mechanism and in the intermediate temperature range, it obeys Greaves' VRH conduction mechanism, thereby suggesting disorder. In addition, the opposite behaviour of the effective paramagnetic magnetic moment and the moment obtained from M vs H as a function of composition; bifurcation of the ZFC and FC magnetic susceptibility data and its decrease with increase in the applied magnetic field suggest strong magnetic anisotropy existing in the compounds under study.

The temperature dependent resistivity of samples 1 and 2 reveal its direct link with the orthorhombic strain. Microscopic understanding of the nature of such strain is vital in unravelling the interactions at play. In the case of sample 4, even though the orthorhombic strain is reduced in comparison to the sample 1, a decrement in the metallicity is observed. This can be understood based on the role played by electron-electron correlation effects and reduction in the number of electrons. These effects lead to the reduction in the density of states (DOS) at the Fermi level that is revealed in the valence band studies. 
The disorder is also expected to localize the electrons at the Fermi level.

We also observe a direct link between the low temperature orthorhombic strain, magnetic anisotropy and the Tc values with composition. The non metallic nature observed in the Cr doped samples also demands careful studies for its utility in the electronic and spintronic devices.

\bibliography{bib}
\bibliographystyle{apsrev4-1}

\end{document}